# Nimesulide limits kainate-induced oxidative damage in the rat hippocampus

Authors:

**Eduardo Candelario-Jalil \*, Hussam H. Ajamieh, Susana Sam, Gregorio Martínez, Olga S. León Fernández**

Affiliation:

Center for Research and Biological Evaluation, Institute of Pharmacy and Food Sciences (CIEB-IFAL), University of Havana, Cuba.

\* Author to whom all correspondence should be addressed

Eduardo Candelario-Jalil, PhD
Apartado Postal 6079
Havana City, 10600
CUBA
Phone: (537)-219531 and 219536
Fax: (537)-336811
E-mail: candelariojalil@yahoo.com

**Abstract**
Kainate induces a marked expression of cyclooxygenase-2 after its systemic administration. Because cyclooxygenase-2 activity is associated to the production of reactive oxygen species, we investigated the effects of nimesulide, a selective cyclooxygenase-2 inhibitor, on kainate-induced in vivo oxidative damage in the rat hippocampus. A clinically relevant dose of nimesulide (6 mg/kg, i.p.) was administered 3 times following kainate application (9 mg/kg, i.p.). After 24 h of kainate administration, the drastic decrease in hippocampal glutathione content and the significant increase in lipid peroxidation were attenuated in nimesulide-treated rats, suggesting that the induction of cyclooxygenase-2 is involved in kainate-mediated free radicals formation.

## 1. INTRODUCTION

Kainate, a pyrrolidine excitotoxin isolated from the seaweed *Digenea simplex,* is a potent neuroexcitatory drug, which after intracerebral or systemic injection leads to generalized limbic seizures in rats (Ben Ari, 1985). Kainate-induced seizures are accompanied by severe neuronal damage predominantly in the hippocampus and amygdala/piriform cortex (Heggli and Malthe-Sörenssen, 1982). The pattern of damage induced by systemically administered kainate approximates to that seen following repeated temporal lobe seizures (Ben Ari, 1985) and cerebral ischemia-reperfusion (Dykens et al., 1987). Therefore, kainate provides a valuable tool with which to model some features of ischemic damage and of the injury induced by repeated epileptic seizures (Sperk et al., 1985).

Current models of kainate toxicity support the hypothesis that the main cause of neurotoxicity is the activation of presynaptic kainate receptors and the release of endogenous glutamate (Ferkany et al., 1982). The overstimulation of glutamate receptors has been implicated in the mediation of injury caused by neurotoxins and ischemia-related insults (Choi and Rothman, 1990). Further, kainate is also thought to mediate damage partly through an indirect mechanism, which may involve the overproduction of reactive oxygen species. The generation of free radicals appears to be pivotal in kainate neurotoxicity (Cheng and Sun, 1994; Carriedo et al., 1998).

It was recently found that kainate induces a marked expression of cerebral cyclooxygenase-2 mRNA and protein following its systemic administration (Hashimoto et al., 1998; Sandhya et al., 1998; Sanz et al., 1997), but the link between kainate-mediated cyclooxygenase-2 induction and free radicals formation has not been clearly defined. In order to investigate the possibility that the induction of cyclooxygenase-2 is involved in kainate neurotoxicity, we examined the effects of nimesulide, a selective cyclooxygenase-2 inhibitor, on kainate-induced in vivo oxidative damage in the rat hippocampus.

## 2. MATERIALS AND METHODS

*2.1. Drugs*
Kainate was purchased from Sigma Chemical Co. (St. Louis, MO, USA). Nimesulide was kindly provided by Gautier-Bagó Laboratories (Buenos Aires, Argentina). The Bioxytech LPO-586 kit for lipid peroxidation was obtained from Oxis International (Portland, OR, USA). All other reagents were of the highest quality available.

*2.2. Animals*
Male S.D. rats (CENPALAB, Havana, Cuba) with a body weight of 200-240 g were used for the experiments. The animals were housed in groups of 4 per cage in a room with controlled 12:12 light/dark cycle and ad libitum food and water.

*2.3. Experimental design*
Four groups of rats ($n$=8 each) were prepared: (1) Kainate (9 mg/kg, i.p., in saline); (2) nimesulide (dissolved in polyvinylpyrrolidone (PVP) in the relation 1:4) was administered (6 mg/kg, i.p.) immediately and after 1 and 2 h following kainate application (as in group 1); (3) saline + nimesulide and (4) saline + PVP (as the control group).

*2.4. Evaluation of behavioral changes*
The behavior of the animals was evaluated during 4 h after injection of kainate according to the following rating scale (Sperk et al., 1985): 0: normal, rare wet dog shakes, no convulsions; 1: intermediate number of wet dog shakes, rare focal convulsions affecting head and extremities, staring and intense immobility; 2: frequent wet dog shakes, frequent focal convulsions (no rearing or salivation); 3: frequent wet dog shakes, progression to more severe convulsions with rearing and salivation (but without falling over); 4: continuous generalized seizures (rearing, falling over), salivation; 5: generalized tonic-clonic seizures, death within 4 h in status epilepticus.

*2.5. Sample preparation*
After 24 h, animals were anesthetized with diethyl ether and perfused intracardially with ice-cold saline. Brains were quickly removed and frozen at -20°C until assay; the hippocampus was dissected on a cold stage before freezing. Hippocampi from treated rats were homogenized in 20 mM Tris-HCl buffer (pH 7.4) and centrifuged for 10 min at 12 000 x g. The supernatant was collected and immediately tested for lipid peroxidation and glutathione content.

*2.6. Lipid peroxidation assay*
Lipid peroxidation was assessed by measuring the concentration of malonaldehyde and 4-hydroxyalkenals using the Bioxytech LPO-586 kit. The assay was conducted according to the manufacturer's instructions.

*2.7. Glutathione (GSH) assay*
Hippocampal glutathione content was measured spectrophotometrically in the deproteinized samples according to the procedure of Ellman (1959).

*2.8. Protein assay*

Total protein concentration was determined using the Coomassie Blue method (Spector, 1978) with bovine serum albumin as standard.

*2.9. Statistical analysis*

Data are expressed as mean values ± SD (standard deviations). Results were analyzed by one-way analysis of variance (ANOVA). If the F values were significant, the Students-Newman-Keuls post-hoc test was used to compare groups. Statistical significance was accepted for P<0.05 and P<0.01.

## 3. RESULTS

*3.1. Behavior*

The most characteristic behavioral changes observed after i.p. injection of kainate (9 mg/kg) were, in order of appearance, strong immobility, increased incidence of wet dog shakes and, initially, focal convulsions. Within 40-60 min after kainate administration the convulsive activity progressed to generalized limbic seizures. Although the intensity of these symptoms showed considerable inter-individual variation, the average symptom rating in both kainate-treated groups reached a value of 3. Nimesulide did not modify the kainate-induced behavioral changes.

*3.2. Effects of nimesulide on kainate-mediated oxidative damage*

After 24 h, the systemic administration of kainate (9 mg/kg, i.p.) produced a drastic decrease in hippocampal GSH levels (0.987 ± 0.06 μmol/g tissue) as compared to those in control group (1.316 ± 0.03 μmol/g tissue; P<0.01). Nimesulide (6 mg/kg, i.p., administered 3 times) partially prevented the dramatic decrease in GSH of kainate-challenged rats (1.182 ± 0.03 μmol/g tissue; P<0.05) as shown in Table 1.

On the other hand, malonaldehyde and 4-hydroxyalkenals, as an index of lipid peroxidation, increased by 75% in kainate-treated group (6.45 ± 0.13 nmol/mg protein) as compared with control group (3.69 ± 0.24 nmol/mg protein; P<0.01). Kainate-mediated lipid peroxidation was attenuated in nimesulide-administered rats as shown in Fig. 1 (4.68 ± 0.09 nmol/mg protein; P<0.01). The administration of nimesulide without kainate did not alter neither GSH nor malonaldehyde and 4-hydroxyalkenals levels.

## 4. DISCUSSION

The ability of kainate to induce oxidative damage has been well documented in the literature (Cheng and Sun, 1994; Carriedo et al., 1998). It is generally accepted that overactivation of excitatory amino acid receptors triggers marked intracellular $Ca^{2+}$ rises and consequent oxygen radical production. Multiple events may be elicited by kainate-induced cytosolic $Ca^{2+}$ accumulation. Activation of calpains may convert xanthine dehydrogenase to xanthine oxidase with the simultaneous formation of free radicals. Further, $Ca^{2+}$-dependent nitric oxide synthase activation may also contribute to kainate-mediated neuronal damage (Cheng and Sun, 1994). Moreover, the release of arachidonic acid following $Ca^{2+}$-triggered phospholipase $A_2$ activation plays an additional role in kainate-induced free radicals formation (Cheng and Sun, 1994). It has been demonstrated that inhibition of arachidonic acid metabolism protects against kainate-induced seizures and neurotoxicity (Baran et al., 1994).

Neuronal cyclooxygenase-2 expression is detected in certain regions of the rat brain, mainly the hippocampus and cortex under physiological conditions (Breder et al., 1995) and can be markedly induced under certain stimuli and insults, including seizures (Adams et al., 1996) and ischemia (Planas et al., 1995; Nogawa et al., 1997). Although the factors responsible for the cytotoxicity of cyclooxygenase-2 have not been clearly defined, it is likely that one of the mechanisms is related to production of reactive oxygen species, which are formed by the peroxidase step of the cyclooxygenase reaction. Cyclooxygenase-2 enzymatic activity can also mediate tissue damage by producing proinflammatory prostanoids (Seibert et al., 1995).

As previously reported, systemic administration of kainate is able to increase prostaglandin formation and this event may be partially responsible for the tissue damage seen after kainate or the consequences of it (Baran et al., 1987).

The effects of nimesulide have not previously been studied using the systemic kainate model of neurotoxicity. Our present results showed that nimesulide, a selective cyclooxygenase-2 inhibitor, partially protected against kainate-induced oxidative damage. These results are consistent with previous reports, which provide strong evidence that cyclooxygenase-2 is implicated in excitotoxicity and neuronal death following brain insults such as ischemia-reperfusion (Nogawa et al., 1997; Nakayama et al., 1998).

In summary, the present study demonstrated that the administration of a clinically relevant dose of the cyclooxygenase-2 inhibitor nimesulide limited kainate-induced oxidative damage in the rat hippocampus. This finding may suggest that cyclooxygenase-2-derived reactive metabolites are involved in kainate excitotoxicity. On the other hand, nimesulide did not modify kainate-induced seizures. This may indicate that the effects of nimesulide are not sufficient to attenuate the kainate-mediated symptoms.

Table 1
Effects of nimesulide (NIM) on depletion of hippocampal glutathione induced by kainate
Values are mean ± SD.

| Treatment | GSH (μmol/g tissue) |
|---|---|
| Control | 1.316 ± 0.033 |
| Kainate | 0.987 ± 0.058[a] |
| Kainate + NIM | 1.182 ± 0.030[b] |
| NIM | 1.314 ± 0.020 |

[a] Significantly different from control ($P < 0.01$).
[b] Significantly different from kainate and control ($P < 0.05$).

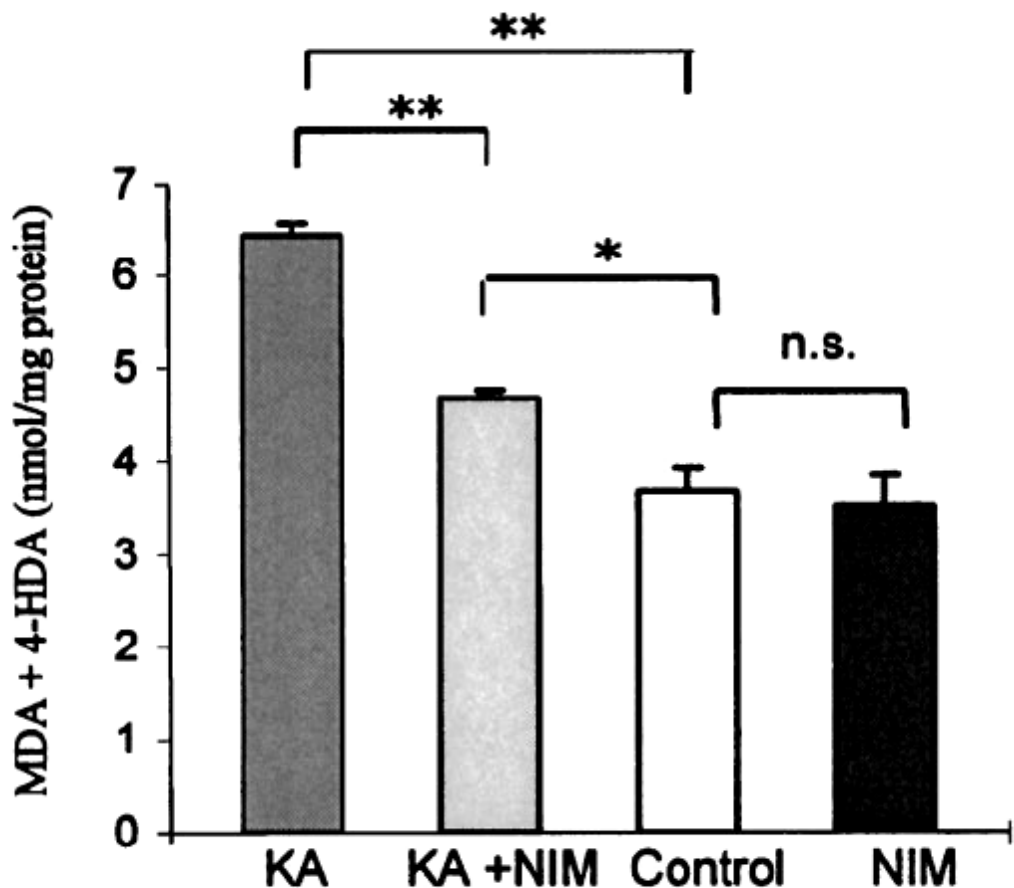

Fig. 1. Effect of kainate (KA, 9 mg/kg, i.p.) and administration of nimesulide (NIM, 6 mg/kg, i.p., three times) on levels of malonaldehyde (MDA) and 4-hydroxyalkenals (4-HDA) in the rat hippocampus. $^*P < 0.05$ and $^{**}P < 0.01$.